\documentclass[12pt,preprint]{aastex}
\slugcomment{Submitted to ApJ Part 1}
\begin{document}
\title{A Scattered Light Echo around SN 1993J in M81}
\author{Ji-Feng Liu, Joel N. Bregman and Patrick Seitzer}
\affil{Astronomy Department, University of Michigan, MI 48109}
\begin{abstract}
A light echo around SN  1993J was observed 8.2 years after explosion by a HST
WFPC2 observation, adding to the small family of supernovae with light echoes.
The light echo was formed by supernova light scattered from a dust sheet, which
lies 220 parsecs away from the supernova, 50 parsecs thick along the line of
sight, as inferred from radius and width of the light echo. The dust inferred
from the light echo surface brightness is 1000 times denser than the intercloud
dust. The graphite to silicate fraction can not be determined by our BVI
photometric measurements, however, a pure graphite model can be excluded based
on comparison with the data.  With future observations, it will be possible to
measure the expansion rate of the light echo, from which an independent
distance to M81 can be obtained.

\end{abstract} 


\section{Introduction}
Shortly after the burst of Nova Persei 1901, a light echo was observed to
emerge with superluminal expansion. It was later properly explained (Couderc
1939) to be the nova light scattered by dust nearby which reached us later than
the unscattered photons due to light travel effects.  The possibility of
observing such an effect with supernovae was later discussed by many authors
(e.g. Zwicky 1940, Schaefer 1987).  More than its splendid appearance, a light
echo also sheds light on the circumstellar and interstellar environments
through which it passes and the most famous example is from SN 1987A in Large
Magelanic Cloud.  Observations of re-emission from the rings around SN 1987A
provide us the information on the geometry, distribution and composition of its
circumstellar medium (e.g., Lundqvist et al 1991) , help us to infer the
stellar wind history of its pregenitor, and even enable us to determine an
accurate distance to SN 1987A (Panagia et al. 1991). Also, the monitoring of
its scattered light echoes enables a three dimenional mapping of the structure
of the interstellar medium in front of SN 1987A (Xu et al. 1995), which reveals
aspects of dense clouds and superbubbles that are difficult to reveal by other
means. 

This phenomenon has also been found in distant galaxies. 
For example, SN 1991T,
a luminous Type Ia supernova in NGC 4527, exhibited a nearly flat light
curve more than 950 days after maximum light, and spectral features that,
although were present in earlier spectra, were substantially narrower and
blueshifted on a significantly bluer continuum (Schmidt et al. 1994).
Schmidt et al. attributed these features to a light echo, which was later 
confirmed by HST FOC observations (Sparks et al.  1999). 
Similar photometric and spectroscopic behaviors in the late-time
observations of SN 1998bu have led to suspicion of a light echo (Cappellaro et
al. 2001), but it has yet to be confirmed by direct high-resoulution imaging.

Due to limited spatial resolution, direct imaging observations of supernova
light echoes are possible only in our local supercluster, which makes these
phenomena rare events.
Thus far, only
Nova Persei 1901, SN1987A, SN1991T and SN1998bu are reported to have associated
light echoes in the literature. Here we report on a light echo around SN1993J,
the fourth such event.

The supernova SN1993J exploded on March 28, 1993 in the spiral galaxy M81,
and due to its proximity (3.6 Mpc), it has been observed and at
wavebands from radio to $\gamma$-ray regions.
It began as a type II supernova, but changed later to type Ib at the
nebular stage, and was classified as type IIb. A series of observations with
the Very Large Baseline Interferometer revealed an expanding radio shell that
was decelerating (Bartel et al.  1999), reflecting the interaction between
the shock front and the circumstellar medium. Intensive photometric and
spectroscopic observations also showed an infrared excess after day 50,
which may be indictive of an infrared echo (Lewis et al. 1994).

In this paper we report an optical scattered light echo around SN 1993J that
was dicovered in an HST WFPC2 observation. In $\S$2, we discuss our
observations showing the light echo, and another archive HST WFPC2 observation
with a nondetection of such a light echo. Models for this light echo are
discussed in $\S$3, which give the geometry and dust properties. For the
distance to SN 1993J, we use the distance to M81, i.e., 3.63 Mpc
($\mu=27.8\pm0.2$, Freedman et al. 1994).

\section{Observations}

As part of an observation of a different target, the WFPC2 was oriented so that
SN1993J fell on a Wide Field chip (WF4) during our 2000 second HST observations
in the B, V, and I bands (F450W, F555W, and F814W).  The observation was
obtained on June 4, 2001, 8.2 years after its explosion; the data were analyzed
following standard procedures.  A light echo is clearly visible around SN1993J,
as shown in Fig. 1.  This ring, which is most obvious in the F555W image, has a
radius of about 1.9'' (19 pixels) and about 0.2'' in width. This ring is a
partial arc with the brightest part about 4.3'' in length. For comparison,
another WFPC2 observation on Jan.  31, 1995 (i.e., 1.8 years after SN1993J)
with supernova centered on the Planetary Camera was also extracted from the
MAST archive.  There was no evidence of similar light echoes or partial arcs in
the 1995.

The mean surface brightness of the arc was computed from those arc pixels that
are brightest in the F555W image, excluded were those pixels that show a faint
source in the F814W image. The background mean and sigma were computed from a
square box centered on the arc, but excluding the arc and point sources.  The
net count rates (in $DN/s/pix^2$) were then converted to surface brightness (in
$mag/arcsec^2$) using zeropoints appropriate for standard BVI photometry.  The
measured surface brightness and lower/upper limits in units of $mag/arcsec^2$
for $1\sigma$ errors in count rates are listed in Table 1. 
For the 1995 observation, the $3\sigma$ detection threshold in the F555W image,
defined by the standard deviation of the mean of a region which includes the
positions of the observed 1.9'' arc and the postulated 0.9'' arc (see section
3), is $\sim25.2mag/arcsec^2$, about 15 times fainter than the observed light
arc.

\begin{figure}
\plotone{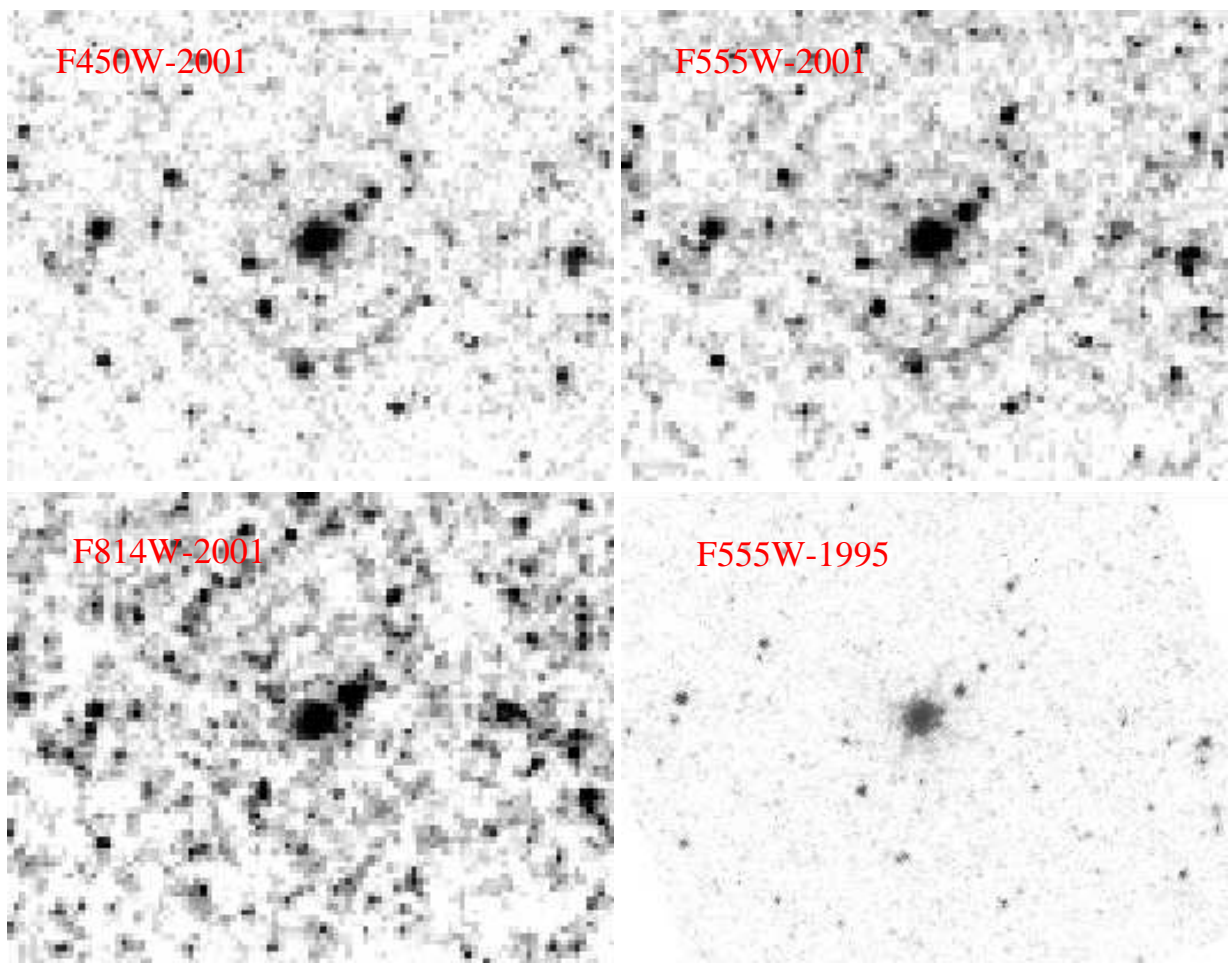}
\caption{A partial arc of a light echo appeared around SN1993J in the HST WFPC2
F450W, F555W and F814W observation from June 2001. The partial arc is brightest
in the F555W and F450W images and has a radius of 1.9'', a width of 0.2'' and
arc length of 4.3''.  No comparable features are seen in
an earlier observation (01/31/1995).}
\end{figure}

\begin{table}
\caption{Surface Brightness of the Light Echo}
\begin{tabular}{|cccc|}
\hline
Filter & brightness & lower limit & upper limit \\
\hline
F450W(B) & 22.61  & 23.75  & 22.49  \\
F555W(V) & 22.26  & 22.35  & 22.18  \\
F814W(I) & 22.46  & 22.80  & 22.20  \\
\hline
\end{tabular}
\end{table}

\section{Discussion}
Scattered light seen time $t$ later after the supernova onset must be scattered
by dust on a $t$-ellipsoid as illustrated in Fig. 2.  For a dust cloud in
front of the supernova, the distance $l$ to the supernova along the line of
sight can be obtained for specified $t$ and the impact parameter $b$ (i.e., the
perpendicular distance from the supernova to the line of sight, $b=D\theta$
where D is the distance to the supernova from the earth, $\theta$ is the angle
of the line of sight from the supernova), from the expression $(l+ct)^2 = b^2 +
l^2$.  For this light echo of SN 1993J, $t=8.2$ years,$\theta=1.9''$, or
$ct=2.5 pc$,$b=33.4 pc$, which leads to, $l\sim220 pc$. This distance, as
well as the asymetric shape of the light echo, indicates that the dust cloud
could not have originated from the red supergiant wind of SN 1993J progenitor.
Instead, it should be a discrete interstellar dust sheet as seen in our galaxy.
The thickness of of the dust sheet along the line of sight can be associated
with the width of the light echo by $\Delta l=(b/ct)\Delta b$.  The observed
values, $\Delta\theta=0.2''$, or $\Delta b\sim3.5 pc$, indicate the dust sheet
has a thickness of about $50 pc$. The geometry described here is illustrated in
Fig.  2. Note, however, that the sheet structure in Fig. 2 is not the only
possible configuration. The detection of this light echo only requires dusty
clouds filling the volumn between the 8.2 year ellipsoid and the
8.2year-120day ellipsoid as shaded in Fig. 2. 

\begin{figure*}
\centering
\plotone{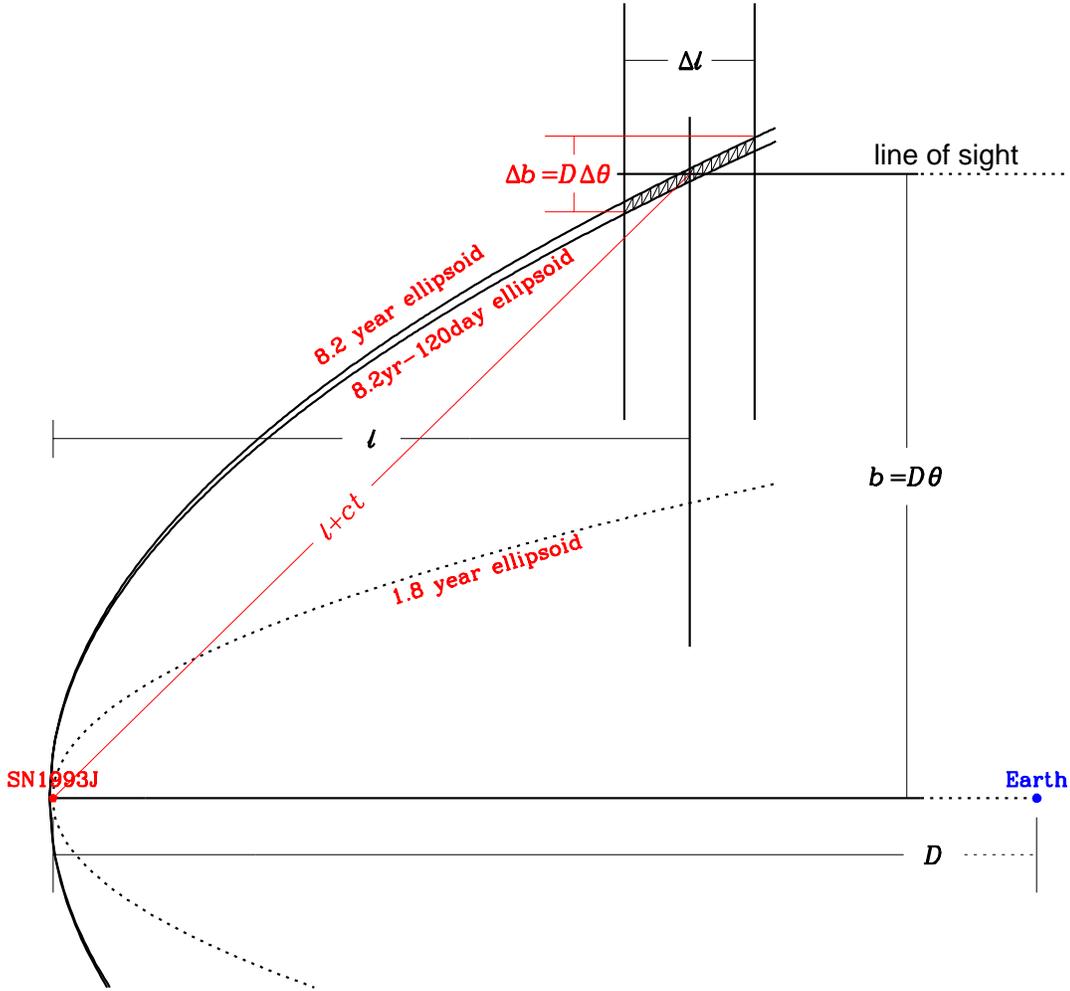}
\caption{Formation of the observed scattered light arc around SN1993J. The 8.2
year ellipsoid is the scattering location for light radiated at supernova
explosion, the 8.2year-120days ellipsoid is for light radiated 120 days after
the supernova explosion. The shaded area between 8.2 year ellipsoid and
8.2year-120day ellipsoid indicate the dust volume that truly contributes to
the scattered light. The symbols are explained in the text. }

\end{figure*}

If this dust sheet extended toward the supernova and crossed the dashed 1.8
year ellipsoid in Fig. 2, a light echo should have been observed by the 1995
WFPC2 observation, with $b\sim15.8 pc$, or $\theta=0.9''$.  The nondetection of
such a light echo indicates that either the sheet of gas and dust does not
extend closer to the supernova or that it becomes thicker and less dense, which
diminishes the surface brightness of the echo. Comparing the detection
threshold of the 1995 observation with the observed surface brightness of the
light echo, we conclude that the dust density of the cloud crossing the 1.8
year ellipsoid must be 10 times less than the region revealed by the light arc.

The cloud extension is also limited by the length of the arc, which indicates
that the dust sheet is at least $60 pc$ wide.

The surface brightness of the scattered light arc is, in units of flux per
square arc second, 

\begin{eqnarray*}
\Sigma_\lambda(b) = & \int\limits_{l_0-\Delta l/2}^{l_0+\Delta l/2}
  \delta[2c(t-\tilde t)l+c^2(t-\tilde t)^2 -b^2]{dl \over \Delta l} \\
 & \times F_{\lambda,0} 10^{-0.4m_\lambda(\tilde t)}{D^2 \over 4\pi 
    r^2 \epsilon^2} N_d  \\
 & \times \int^{a_2}_{a_1}f(a)da\pi a^2Q_{\lambda,sca} F_\lambda(\alpha) 
\end{eqnarray*}

in the above equation, $F_{\lambda,0}$ is the flux of magnitude 0, $l_0$ and
$\Delta l$ are the location and thickness of the dust sheet as explained in the
above subsection, $\epsilon$ is 206265, $N_d$ is the dust column density in
unit of $grain/cm^2$, $a$ is the radius of the grain, $f(a)$ is the normalized
grain size distribution. $Q_{\lambda,sca}$ is the scattering efficiency of the
dust.  $F_\lambda(\alpha)$ is the phase function as in Schaefer (1987) where
$\alpha$ is the scattering angle. The quantity $\tilde t$ is the elapsed time
since supernova explosion,  $m_\lambda(\tilde t)$ is the observed un-dereddened
light curve, which we took the UBVRI photometry of the first 120 days (c.f.,
Lewis et al.  1994). The $\delta$ function in the integral determines which
part of the light curve is scattered by dust at location $l$.

The existence of interstellar dust has long been established from, for example,
the interstellar extinction curve. Fitting to the interstellar extinction curve
gives the composition and size distribution when combined with other
constraints such as cosmic abundances, gas-to-dust ratios, interstellar
physical conditions, and optical properties of the grains. Mathis et al. (1977)
conclude that the simplest description that fits the observations and the
general constraints is of a mixture of uncoated graphite and silicate grains,
with a common power law size distribution $f(a)da\propto a^{-3.5}da$. This
description seems to be valid over the entire sky, giving rise to the
approximately uniform interstellar extinction laws. Local variations from this
composition do exist. For example, Bromage \& Nandy (1984) show that the
extinction curve toward SMC does not show the $2200\AA$ bump that are due
to small graphite grains, and conclude that the usual graphite contribution is
at least a factor of seven weaker than the 'normal' galactic dust.  In our
calculation, we considered models of different mixture of silicate and
graphite, with size limits taken as $0.01-0.25\mu m$ for silicate grains and
$0.005-0.25\mu m$ for graphite grains (Mathis, Rumpl, and Nordsieck, 1977). The
grain properties are taken from Draine's calculations (Draine et al. 1984).  

\begin{figure}
\plottwo{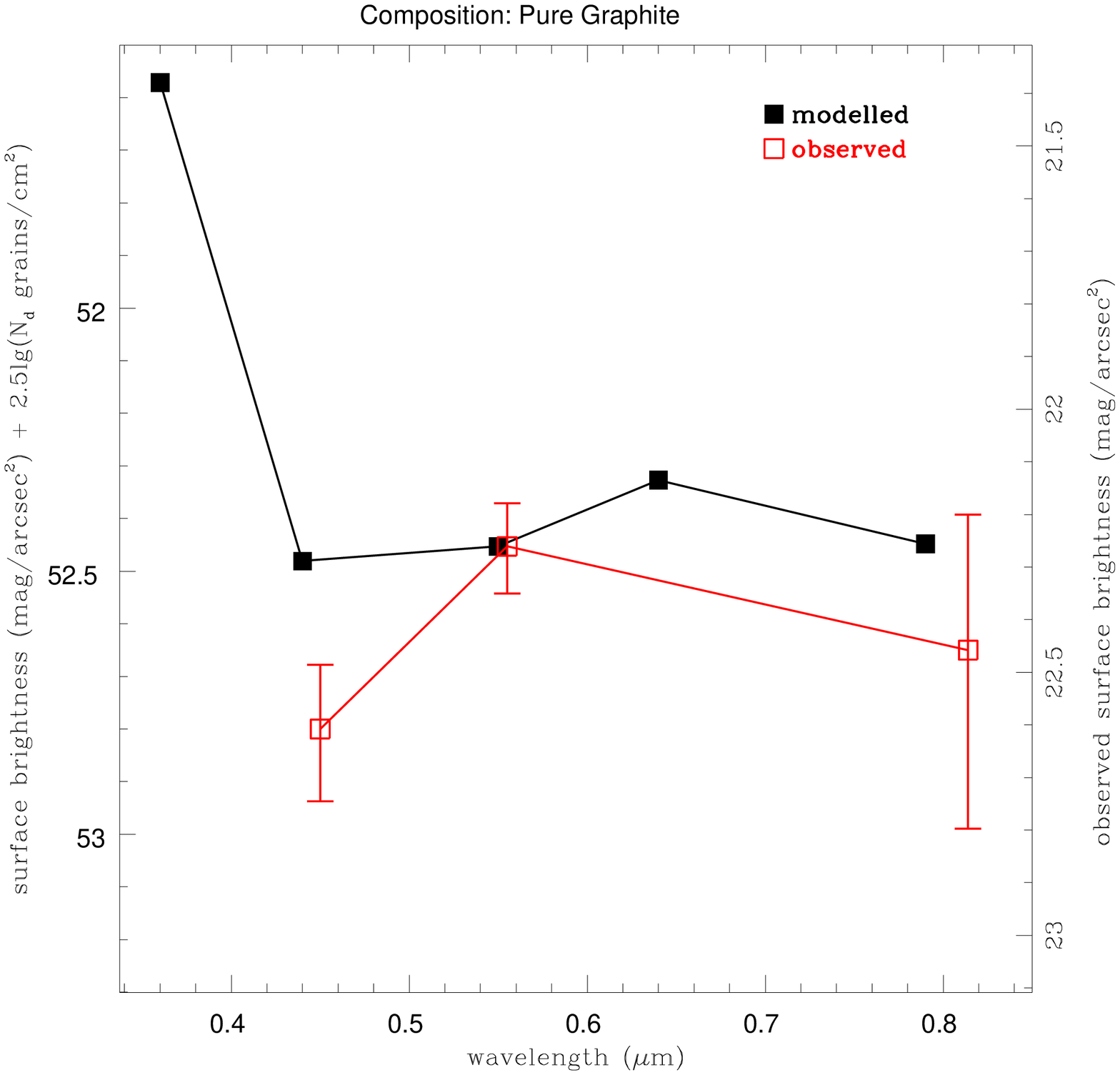}{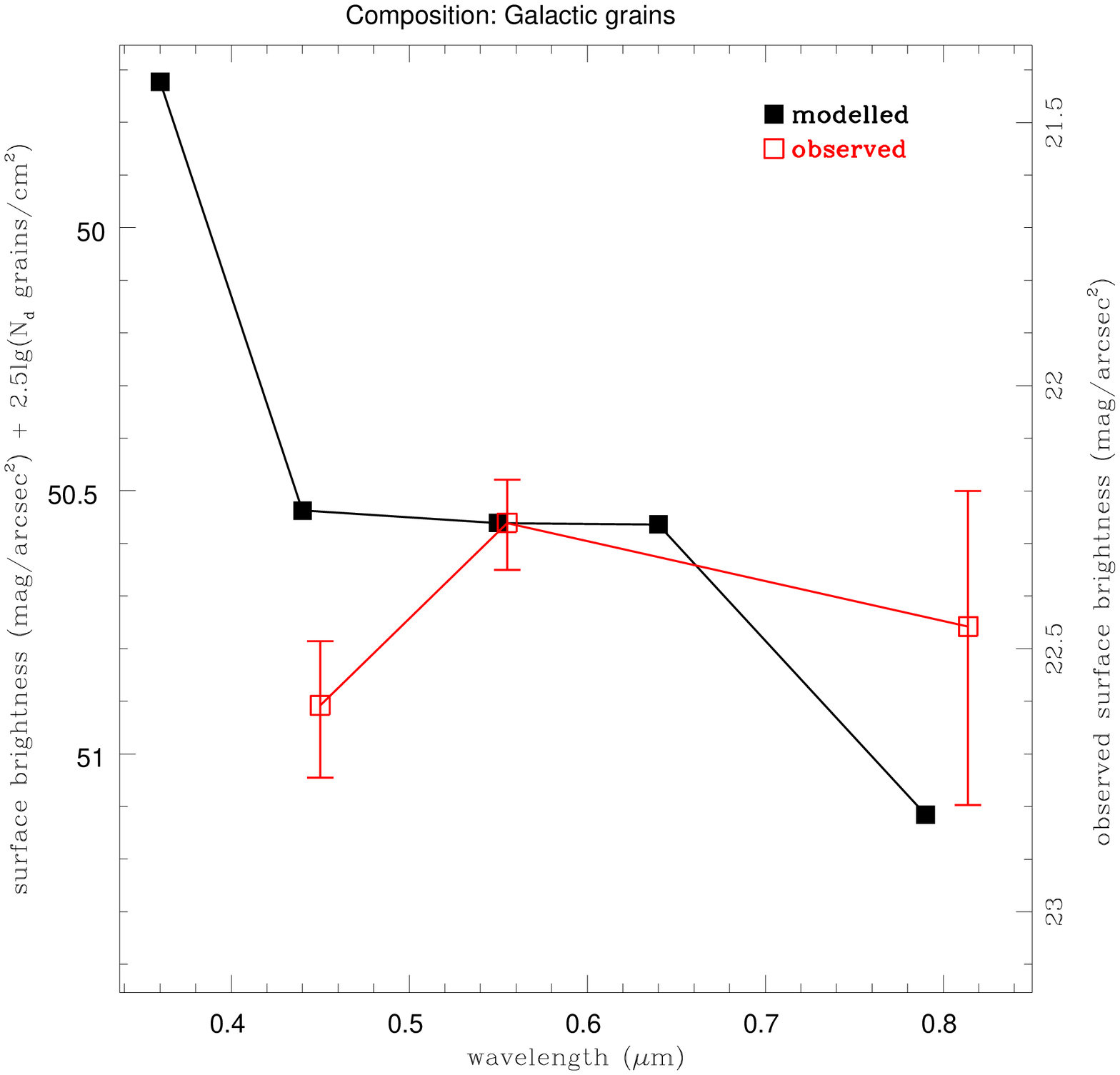}
\end{figure}
\begin{figure}
\plottwo{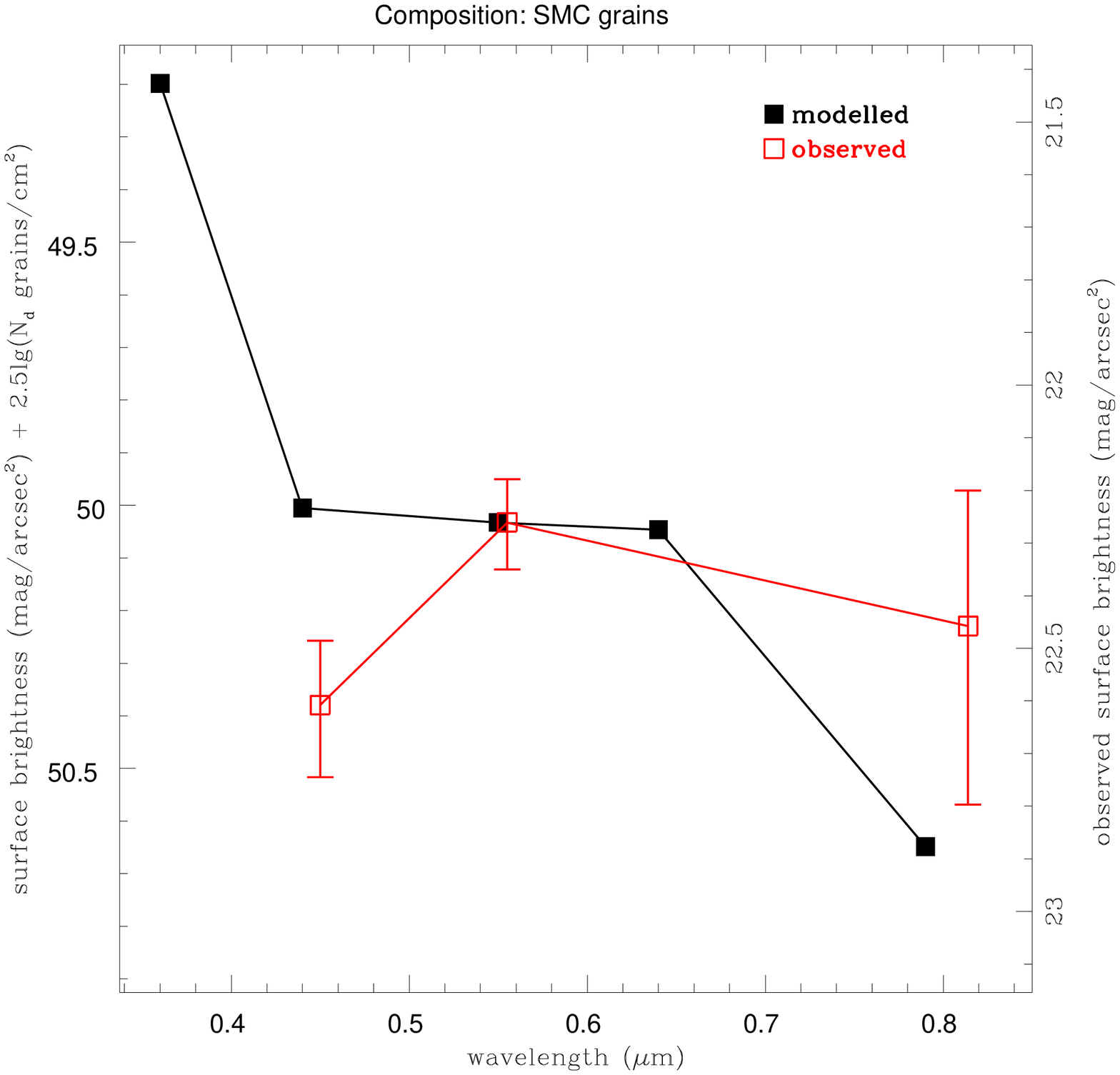}{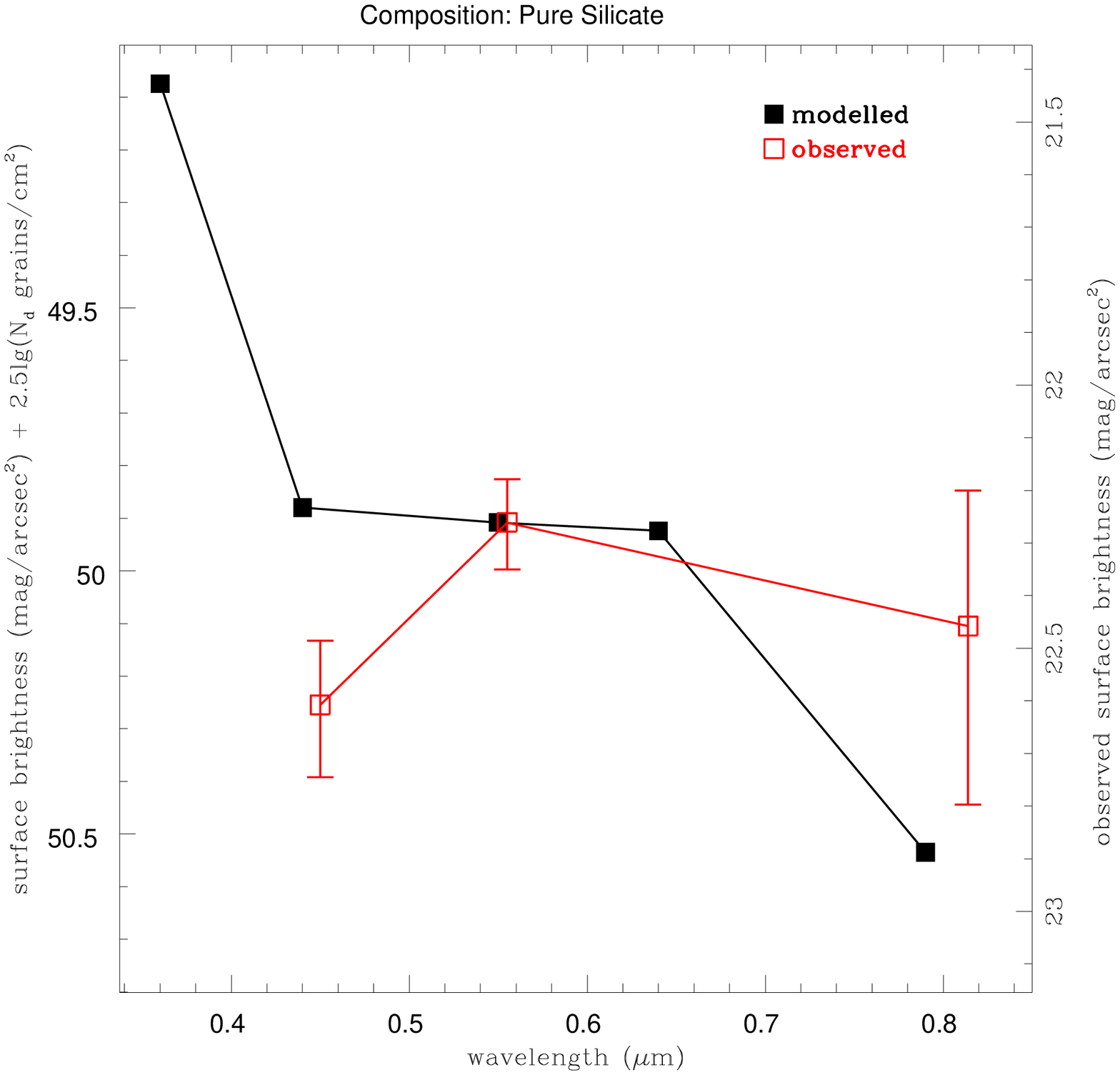}
\caption{Observed and calculated surface brightness of the light echo. The four
models are pure graphite model, SMC dust model, normal galactic dust model, and
pure silicate model. The dip in the calculated curve around B band is due to
the fact that B band light of SN 1993J is much fainter than the redder bands.}
\end{figure}

The observed surface brightness and the calculated values in UBVRI
for four grain models are shown in Fig. 3. By comparing with the observed
surface brightness, the grain column density can be obtained. Although the
scattering effeciency is greater for blue light, the scattered light in B band
is no brighter than V and R band, due to the fact that the supernova light in B
band is much fainter than in redder bands. In our pure graphite model (model
A), the scattered light in R and I band is much brighter than in B and V band.
This feature, however, can not be used to exclude the pure graphite model due
to the large measurement uncertainties.  
The other models, the Galactic
grain model (model B), the SMC grain model (model C) and the pure silicate
model (model D), are consistent with the data.  For a pure silicate model,
comparing measured V band surface brightness with the model, a grain column
density $\sim1.1\times10^{11}grains/cm^2$ can be infered.Given that the dust
sheet is about $50 pc$ wide, this leads to a grain density of
$\sim7.4\times10^{-10}grains/cm^{-3}$, about 1000 times denser than the
intercloud density ($0.5\times10^{-12} grains/cm^{-3}$;Allen 1976). This
density increases when more graphite is introduced into the model. For the
Galactic grain model, the inferred grain density is
$\sim1.4\times10^{-9}grains/cm^{-3}$, which corresponds to a Hydrogen column
density $3.9\times10^{20} cm^{-2}$ if we adopt an empirical relation,
$N_H=5.8\times10^{21}*E(B-V)$ (Bohlin et al. 1979) .

Further constraints may be placed when we consider the extinction of star light
due to these dust, which can be obtained by $$A_\lambda =
1.086N_d\int^{a_2}_{a_1} \pi a^2 Q_{\lambda,abs}(a) f(a) da$$ The extinctions
in standard UBVRI bands are listed in table 2 for the four models.  The pure
graphite model can be rejected by this extinction argument, since a grain
column density of $1.2\times10^{12} grains/cm^2$ is needed to produce the
observed surface brightness and will cause an extinction $A_V\sim1.6 mag$,
which is much larger than the observed extinction of SN 1993J of $A_V\sim0.25
mag$ (Richmond et al.  1994), and would have blocked all background stars in
the 1995 observation.

\begin{table}
\caption{Extinction (mag) for four grain models}
\begin{tabular}{|cccccc|} 
\hline 
Model & $A_U$ & $A_B$ & $A_V$ & $A_R$ & $A_I$ \\
\hline 
A  & 2.54   & 2.12   & 1.68   & 1.43   & 1.12  \\
B  & 0.37   & 0.30   & 0.23   & 0.20   & 0.15  \\
C  & 0.19   & 0.15   & 0.11   & 0.09   & 0.07  \\
D  & 0.16   & 0.13   & 0.09   & 0.08   & 0.06  \\
\hline
\end{tabular}
\end{table}

\section{Conclusions}
We have presented the WFPC2 observation of a light echo around SN 1993J after
8.2 years of supernova explosion. This light echo is due to light scattered
from a dust cloud about $220 pc$ in front of SN 1993J, about $50 pc$ thick and
$60 pc$ wide, revealing the existence of a sheet of dust (and gas) in another
galaxy.  The dust inferred from the light echo surface brightness is 1000 times
denser than the intercloud dust. The graphite to silicate fraction can not be
determined by our BVI photometric measurements, however, a pure graphite model
can be excluded based on comparison with the data.

Aside from studying the geometric structure of the interstellar medium in other
galaxies, a light echo can be used to determine the distance to the host galaxy.
To accomplish this, one needs to measure the expansion rate of the light echo.
With future observations, we should be able to determine this expansion rate
and obtain a measurement of the distance independent of that obtained by using
Cepheid variables.

\acknowledgements

we are grateful for the service of MAST. JL thanks Louis-Gregory Strolger for
helpful discussions. We gratefully acknowledge support for this work from NASA,
grant HST-GO-09073.01.

\end{document}